\documentclass[showpacs,twocolumn,showkeys,amsmath,amssymb,pra,nofootinbib,subscriptaddress,longbibliography
]{revtex4-1}

\usepackage{amsmath,amssymb,amsfonts}
\usepackage{graphicx}
\usepackage{txfonts}
\usepackage{epstopdf}
\usepackage[T1]{fontenc}

\usepackage[unicode]{hyperref}
\hypersetup{
   unicode=true,          
   plainpages=false,
   pdftitle={Title of PDF},    
   pdfauthor={Author of PDF},     
   pdfsubject={Subject of PDF},   
   colorlinks=true,       
}
\newcommand{\Zce}[0]{\mbox{Z}}
\newcommand{\Zgce}[0]{\Xi}

\begin{document}

\title{Grand-canonical condensate fluctuations in weakly interacting
  Bose-Einstein condensates of light
}

\author{Christoph Weiss}
\email{christoph.weiss@durham.ac.uk}
  \affiliation{Joint Quantum Centre (JQC) Durham--Newcastle, Department of Physics, Durham University, Durham DH1 3LE, United Kingdom}

\author{Jacques Tempere}
\affiliation{TQC--Theory of Quantum and Complex Systems, Universiteit Antwerpen, Universiteitsplein 1, B-2610 Antwerpen, Belgium}

\date {13 September 2016}

 \begin{abstract}
Grand-canonical fluctuations of Bose-Einstein condensates of light are
accessible to state-of-the-art experiments [J.\ Schmitt \textit{et
  al.}, Phys.\ Rev.\ Lett.\ \textbf{112}, 030401 (2014).].
We
phenomenologically describe these fluctuations by using the
grand-canonical ensemble for a weakly interacting Bose gas at thermal
equilibrium. For a two-dimensional harmonic trap, we use two models
for which the canonical partition functions of the weakly interacting
Bose gas are given by exact recurrence relations. We find that the
grand-canonical condensate fluctuations for weakly interacting Bose
gases vanish at zero temperature, thus behaving qualitatively similar
to an ideal gas in the canonical ensemble (or
micro-canonical ensemble) rather than the
grand-canonical ensemble. For low but finite temperatures, the fluctuations remain
considerably higher than for the canonical ensemble, as predicted by
the ideal gas in  the grand-canonical ensemble, thus clearly showing that we are not in a regime in which the ensembles are equivalent. 
\end{abstract}


\maketitle

\section{Introduction}

The experimental realization of Bose-Einstein condensates (BECs) of
photons~\cite{KlaersEtAl2012,MarelicNyman2015} opens new areas of
research
 beyond Bose-Einstein condensates in ultracold atomic
 gases~\cite{BlochEtAl2008}: {Theoretical investigations of
 photon condensates~\cite{Sobyanin2013,KirtonKeeling2013,deLeeuwEtAl2014,KirtonKeeling2015,ChiocchettaEtAl2015} are supplemented by research on quantum phase transitions of light~\cite{GreentreeEtAl2006}, polariton condensates~\cite{CristofoliniEtAl2013,BaliliEtAl2007} and the observation of kinetic condensation in classical waves \cite{SunEtAl2012}. }
While in experimental realisations of BECs with ultracold
atoms~\cite{AndersonEtAl1995,DavisEtAl1995,BradleyEtAl1995} the total number of atoms is (approximately) constant,
for photon-BECs they vary enormously: the experiment of
Ref.~\cite{SchmittEtAl2014} shows grand-canonical statistics close to
the ideal (non-interacting) gas. 

Here, we use the grand-canonical ensemble to describe such a BEC in
the \textit{presence of weak}\/ interactions.
Non-interacting bosons have been used to gain
insights into number-theoretical questions~\cite{SchumayerHutchinson2011,Riemann} like the number
partitioning
problem, the question how many ways can an integer be expressed as sums of integers (for which deviations from Gaussian fluctuations have been investigated by applying methods developed for ultracold atoms to number
 theory both for bosons~\cite{WeissHolthaus2002,WeissEtAl2003} and fermions~\cite{KubasiakEtAl2005}). A related question is how many possible ways there are to express a number as products of numbers~\cite{WeissEtAl2004b} ---
here bosons in a logarithmic potential can be used~\cite{GleisbergEtAl2013,GleisbergEtAl2015}.

For such number-theoretical problems, non-interacting particles are
the obvious choice. However, for current experiments ideal Bose gases can
only be an approximation. This would of course not be an issue if real gases always approached the ideal gas provided the interaction is week enough. However, as we will see for low temperatures, even very weak interactions that hardly change the number or particles in the condensate can change the ideal-gas predictions for grand-canonical condensate fluctuations considerably. 

Thus,
the research on ideal Bose gases using the canonical ensemble or the
micro canonical ensemble, for example, for fluctuations of ideal gases arguably are particularly useful if it survives weak interactions.  Anomalous condensate fluctuations --- condensate fluctuations that scale faster with particle number $N$ than $N^{0.5}$ --- are present in for some (but not all) cases for ideal Bose gases:  only grand-canonical fluctuations become (for ultra-cold atoms unphysically) large for the three-dimensional trap investigated by~\cite{Politzer1996,GajdaRzazewski1997,NavezEtAl1997} whereas the one-dimensional trap used by~\cite{GrossmannHolthaus1996} does yield large fluctuations even in the canonical ensemble. A systematic investigation for which dimensions and trap geometries anomalous fluctuations occur can be found  for canonical ensemble in~\cite{WeissWilkens1997} and for the microcanonical ensemble in~\cite{GrossmannHolthaus1997a}.  

When investigating interacting Bose gases~\cite{SvidzinskyScully2006,KocharovskyEtAl2006,HellerStrunz2013}, the question whether or not anomalous fluctuations survive interactions has been an ongoing debate for quite some time. Weakly interacting Bose-gases have been reported to show anomalous fluctuations if they are confined by a three-dimensional box potential~\cite{GiorginiEtAl1998}, a system for which other authors have investigated the transition from anomalous to normal fluctuations~\cite{IdziaszekEtAl1999}. Even for strongly interacting superfluids anomalous fluctuations have been predicted~\cite{MeierZwerger2001} and explained on a fundamental level~\cite{Zwerger2004}. However, fundamental critisism of anomalous fluctuations has been described in Ref.~\cite{Yukalov2005,Yukalov2005b}: in the thermodynamic limit anomalous fluctuations lead to diverging behavior of quantities important for stability considerations;  the use of a second-order theory for
 calculating fourth-order quantities has also been criticized~\cite{Yukalov2005,Yukalov2005b}.
The divergences criticized in Ref.~\cite{Yukalov2005,Yukalov2005b} only happen in the thermodynamic limit and they thus do not make statements on behaviour for the finite particle numbers on which this paper (and many previous literature) focuses; here fluctuations can still be very large.

When deciding for an ensemble to describe a photonic BEC with, the
\textit{grand canonical ensemble}\/ is the most obvious choice~\cite{SchmittEtAl2014}. While
for an ideal Bose gas there are differences between canonical (or
micro-canonical) and grand-canonical predictions for many thermodnamic
quantities (cf.~\cite{HerzogOlshanii1997,TarasovEtAl2015}), they are
particularly large for fluctuations of the number of particles in the
ground state
(cf.~\cite{GajdaRzazewski1997,GrossmannHolthaus1997a,WeissWilkens1997}). The
grand-canonical rms fluctuations $\Delta n_0$ of the ground state
occupancy of an ideal Bose gas become unphysically 
large when one tries to describe atomic gases (at temperature $T=0$ they 
are larger than the total number of atoms, $N$). However, for photon BECs
such high fluctuations can still be
physical~\cite{SchmittEtAl2014}.  In this paper we investigate the
influence of weak interactions on photon BECs when treated within the
grand-canonical ensemble with two models that can be treated via exact
recurrence relations on the level of the canonical
ensemble
(Refs.~\cite{BrosensEtAl1997a,BrosensEtAl1997b,TempereDevreese1997}
and \cite{IdziaszekEtAl1999}, respectively).

As such, the goal of the present paper is to look specifically at the influence of
\textit{interactions} on the grand-canonical fluctuations. In the grand-canonical
description, it is assumed that the size of the reservoir is substantially larger than
that of the subsystem. 
However, in photonic BECs, 
the size of the reservoir can be changed and made small, so that the cross-over from a
grand-canonical ensemble towards a more canonical ensemble can be investigated. The
study of this cross-over
falls outside the focus of the present paper, in which we want to isolate the effects
of weak 
interactions from effects of reservoir size. Recent work \cite{vanderWurffEtAl2014} 
has suggested that 
weak interactions do affect the grand-canonical predictions in a manner similar to that
observed experimentally in Ref.~\cite{SchmittEtAl2014}. Rather than treating the
interactions variationally as in Ref.~\cite{vanderWurffEtAl2014}, we will use model
systems that allow to take interactions into account in an exact way.

The paper is organized as follows: in Sec.~\ref{sec:calculating} we discuss interacting many-particle 
model systems for which the canonical partition functions are given by exact recurrence relations. We 
introduce the two models used in this paper in Sec.~\ref{sub:modelCW} and  Sec.~\ref{sub:modelJT}. We 
present numerical and analytical results in Sec.~\ref{sec:results} before the paper ends with the 
conclusions in Sec.~\ref{sec:conclusions}.

\section{\label{sec:calculating}Calculating the Grand-canonical partition function using
  exact recurrence relations for the canonical partition functions}
\subsection{Overview of Sec.~\ref{sec:calculating}}
We start this section with introducing the \textit{grand-canonical}\/ ensemble
we use to calculate the thermodynamics of a photon BEC
(Sec.~\ref{sub:grand}). As these equations are based on the \textit{canonical}\/
partition functions, we then discuss the two models we use to
calculate the canonical partition functions via exact recurrence
relations. The model introduced in Sec.~\ref{sub:modelCW} was previously used
in~\cite{IdziaszekEtAl1999}, the model of Sec.~\ref{sub:modelJT}  can
be found in~\cite{TempereDevreese1997}. Both describe an interacting
gas in a two-dimensional harmonic oscillator with $N$ atoms at finite
temperature $T$ phenomenologically. When used to calculate
grand-canonical partition functions, the model of
Sec.~\ref{sub:modelCW}  converges for repulsive interactions, the
model of Sec.~\ref{sub:modelJT}  converges for attractive interactions. Both models are independent of each other, which helps to prevent making model-dependent statements. Both have the advantage of allowing to treat a weakly interacting Bose gas in the grand-canonical ensemble using numerically exact recurrence relations.  For a very weakly interacting Bose gas with pairwise interactions, the model introduced in Sec.~\ref{sub:modelCW} becomes exact for very low temperatures.

\subsection{\label{sub:grand}Grand-canonical ensemble: partition function and
 fluctuations of the total number of bosons}

We will first calculate the canonical partition functions for $N$ particles $\Zce_N(\beta)$ where
\begin{equation}
\beta \equiv \frac 1{k_{\rm B}T}
\end{equation}
and then proceed to calculate the grand-canonical partition function via
\begin{align}
\Zgce(\beta,z) &= \sum_{n=0}^{\infty}z^n\Zce_{n}(\beta)\nonumber\\
&\simeq  \sum_{n=0}^{N_{\rm max}}z^n\Zce_{n}(\beta)
\label{eq:ZGCE}
\end{align}
where we make sure that the sum has already converged at $n\approx
N_{\rm max}$. As we will see later, this puts a constraint on the
sign of the interaction we can model: one of our models can only treat
attractive interactions, the other only repulsive interactions.

The value for $z$ is determined by asking the model to describe an experimental situation with an average number of particles of $N = \langle n \rangle$, where
\begin{equation}
\langle n \rangle=\frac{\sum_{n=0}^{N_{\rm max}}nz^n\Zce_{n}(\beta)}{\sum_{n=0}^{N_{\rm max}}z^n\Zce_{n}(\beta)}.
\end{equation}
Once $z$ is determined (e.g.\ via bisection), the rms fluctuations of the total number of atoms,
\begin{equation}
\Delta N = \sqrt{\langle n^2 \rangle - \langle n \rangle^2}
\end{equation}
 also are accessible via
\begin{equation}
\langle n^2 \rangle=\frac{\sum_{n=0}^{N_{\rm max}}n^2z^nZ_{n}(\beta)}{\sum_{n=0}^{N_{\rm max}}z^n\Zce_{n}(\beta)}.
\end{equation}
Here, and further on, we use lowercase $n$ to indicate the fluctuating 
total number of particles in the open subsystem in the grand canonical 
ensemble. Capital $N$ and $\Delta N$ denote the grand-canonical average and standard
deviation of that number, respectively. When used in the context of the canonical
ensemble, 
$N$ denotes the fixed number of particles in that ensemble. Within the canonical
ensemble, $\Delta N =0$. The number of condensate atoms is noted as $n_0$.

\subsection{\label{sub:modelCW}LTI model system}
The Low Temperature Interaction (LTI) model is based on~\cite{IdziaszekEtAl1999}.

For  very weak interactions and \textit{very}\/ low temperatures we have
\begin{align}
n_0 \simeq N,\\
N_{\rm ex} \ll N
\end{align}
where
\begin{equation}
N_{\rm ex} \equiv N-n_0.
\end{equation}
In this limit, we can 
treat the interaction between ultracold bosons in a harmonic trap analytically.
Setting the single-particle ground state energy to zero, the ground
state energy is proportional to the number of pairs $n_0(n_0-1)/2$ and the
energy for one pair $2\alpha$\footnote{\label{foot:attractiveBEC}For attractive interactions, BECs can become unstable especially in spherical geometries~\cite{DoddEtAl1996,KhaykovichEtAl2002,StreckerEtAl2002}. State-of-the-art experiments show that attractively interacting atomic BECs can nevertheless be stable on experimentally relevant timescales in quasi-one-dimensional wave guides~\cite{KhaykovichEtAl2002,StreckerEtAl2002,CornishEtAl2006,McDonaldEtAl2014,MedleyEtAl2014,NguyenEtAl2014,EverittEtAl2015,MarchantEtAl2016,LepoutreEtAl2016}. While equivalent situations might be difficult to realise with photon BECs, as explained below Eq.~(\ref{n0kgce2}), the reason why we focus on positive interactions for the LTI model is not a stability question: the grand-canonical partition function diverges. If there should be future experiments with attractively interacting photon condensates (for which there currently does not seem to be experimental evidence), this would require a more detailed modeling closer to the experiment.}
\begin{align}
\label{eq:E0}
E_0(N) &= \alpha N(N-1),\\\nonumber\alpha&\ge 0.
\end{align}
The fact that we are in a Bose-condensed state with a macroscopically occupied single particle state is reflected in the energy scaling quadratically (rather than linearly) with particle number $N$.  This scaling of the interaction energy can be found, for example, in the mean-field treatment of Bose-Einstein condensates via the Gross-Pitaevskii equation~\cite{PethickSmith2008}

The total interaction energy is then approximately given by
\begin{align}
\label{eq:E1}
E_{\rm int}^{(1)}(n_0,N) &= \alpha n_0(n_0-1).
\end{align}
Following~\cite{IdziaszekEtAl1999}, we also include the interaction
between condensed atoms and non-condensed atoms
\begin{align}
\label{eq:E2}
E_{\rm int}^{(2)}(n_0,N) &= \alpha n_0(n_0-1)+2\alpha n_0N_{\rm
  ex}\nonumber \\
&= \alpha (Nn_0-n_0^2-n_0).
\end{align}
For the purpose of our grand-canonical calculations, we even can take
a further step and treat interaction between excited atoms in the same
way, yielding
\begin{align}
\label{eq:E3}
E_{\rm int}^{(3}(n_0,N) &= \alpha N(N-1).
\end{align}
While Eq.~(\ref{eq:E1}) underestimates the total interaction energy,
Eq.~(\ref{eq:E3}) overestimates it for finite temperatures. In the
limit of extremely low temperatures all three energies coincide with
the exact interaction energy if the interaction is pairwise as $n_0=N$ at $T=0$ for weak interactions:
\begin{align}
\label{eq:Eexact}
E_{\rm int}^{(j)}(n_0) &= E_{\rm int}^{\rm exact}, \quad k_B T\ll
\hbar\omega, \quad j\in\{1,2,3\}.
\end{align}

For the excited atoms, that is the non-condensed parts, we can then use the recurrence relation~\cite{Landsberg1961}
\begin{align}
\label{eq:rec}
\Zce^{\rm (ex)}_N(\beta)&= \frac 1N \sum_{n=1}^{N}Z^{\rm
  (ex)}_1(n\beta)\Zce^{\rm (ex)}_{N-n}(\beta), \\ Z_0^{\rm
  (ex)}(\beta)&\equiv 1;
\end{align}
for~a~2D~harmonic~oscillator~we~have
\begin{align}
\Zce^{\rm (ex)}_1(\beta) &= \left(\frac1{1-\exp(-\beta\hbar\omega)}\right)^2-1
\end{align}
for the canonical partition functions.
Derivations of the recurrence relation~(\ref{eq:rec}) can be found in Refs.~\cite{BorrmannFranke1993,BrosensEtAl1996,WeissWilkens1997,ChaseEtAl1999}.

The total canonical partition function is then given by
\begin{align}
\label{eq:Zcetotal}
\Zce^{(j)}_N(\beta) &= \sum_{n_0=0}^N \exp\left[-\beta E_{\rm int}^{(j)}(n_0,N)\right]\Zce^{\rm (ex)}_{N-n_0}(\beta), \quad N\ge 1\\
\Zce_0(\beta) &\equiv 1.
\end{align}
The grand-canonical partition function then reads
\begin{align}
  \label{eq:Zgcetotal}
\Zgce^{(j)}(\beta,z) &= \sum_{n=0}^{\infty}z^n\Zce^{(j)}_{n}(\beta)
\end{align}

The canonical expectation for the average occupancy ($k=0$) and higher moments
($k>1$)  are given by
\begin{equation}
 \langle n_0^k\rangle_{N,j}^{\rm can} 
= 
\frac 1{\Zce^{(j)}_N(\beta)}\sum_{n_0=0}^N n_0^k
\exp\left[-\beta E_{\rm int}^{(j)}(n_0,N)\right]\Zce^{\rm (ex)}_{N-n_0}(\beta).
\label{eq:ceav}
\end{equation}
So far, this section has focused on the canonical ensemble with fixed total particle number $N$ within the LTI model. In order to obtain the grand-canonical counterpart of Eq.~(\ref{eq:ceav}), we have to add a grand-canonical average to obtain:
\begin{equation}
\label{n0kgce2}
 \langle n_0^k\rangle^{\rm gc}_j 
=
\frac 1{\Zgce^{(j)}(\beta)}\sum_{n=0}^{\infty}\langle n_0^k\rangle_{n,j}^{\rm can}  z^n \Zce^{(j)}_n(\beta).
\end{equation}

While for attractive \textit{atomic}\/ BECs stability is an issue (see footnote~\ref{foot:attractiveBEC}), for a (so far hypothetical) grand-canonical \textit{photon}\/ BEC with attractive pairwise interaction the grand-canonical partition function would diverge [see Eqs.~(\ref{eq:Zcetotal}) and (\ref{eq:Zgcetotal})] because of contributions of arbitrarily high photon number, thus indicating the necessity to either rethink the modeling of such a system or restrict the experiments to a regime in which the grand-canonical ensemble is not valid.


\subsection{\label{sub:modelJT}BDL model system~\cite{BrosensEtAl1997a}}

The model system of Brosens, Devreese and Lemmens\cite{BrosensEtAl1997a,WeissTempere2016Sup} 
(hereafter referred to as BDL) consists of $N$
particles in a harmonic trap, with harmonic interactions between each pair of
particles:%
\begin{equation}
L\left(  r_{j},\dot{r}_{j}\right)  =\sum_{j=1}^{N}\left(  \frac{m}{2}\dot
{r}_{j}^{2}-\frac{m\Omega^{2}}{2}r_{j}^{2}\right)  +\frac{m\omega^{2}}{2}%
\sum_{j,\ell>j}^{N}\left(  \mathbf{r}_{j}-\mathbf{r}_{\ell}\right)  ^{2}.
\end{equation}
Introducing the center-off-mass and relative coordinates%
\begin{align}
\mathbf{R} &  =\frac{1}{N}\sum_{j=1}^{N}\mathbf{r}_{j},\\
\mathbf{x}_{j} &  =\mathbf{R}-\mathbf{r}_{j},
\end{align}
the Lagrangian can be rewritten as
\begin{align}
L\left(  x_{j},\dot{x}_{j};R,\dot{R}\right)  =~&\frac{ N m }%
{2}\dot{R}^{2}-\frac{N m \Omega^{2}}{2}%
R^{2}\nonumber\\
&+\sum_{j=1}^{N}\left(  \frac{m}{2}\dot{x}_{j}^{2}-\frac{m\left(
\Omega^{2}+N\omega^{2}\right)  }{2}x_{j}^{2}\right)  .
\end{align}
This new Lagrangian is of course subject to the constraint $\mathbf{R}%
=\frac{1}{N}\sum_{j=1}^{N}\mathbf{r}_{j}$, so although it looks as if it has
$N+1$ degrees of freedom per spatial dimension, there are only $N$ independent
ones. Of these, one has eigenfrequency $\Omega$ and $N-1$ have 
eigenfrequency
\begin{equation}
w=\sqrt{\Omega^{2}+N\omega^{2}}.\label{spectrum}%
\end{equation}
Switching the sign of the last term in the Lagrangian turns the interactions
from attractive to repulsive, and leads to $w=\sqrt{\Omega^{2}-N\omega^{2}}$. This 
means that for repulsive harmonic interactions, there is an upper bound to the number of particles that can be held together by the trap. In order for the grand-canonical ensemble to include all number states we therefore restrict ourselves to the case of attractive interactions. In contrast to the attractive contact potential, for a harmonic potential this does not lead to a collapse, but rather to a change of the trapping frequency $\Omega$ to $w$ for $N-1$ (D-dimensional) degrees of freedom. The remaining center of mass degrees of freedom acquire a frequency $\Omega$.
In a series of 
papers\cite{BrosensEtAl1997a,BrosensEtAl1997b,BrosensEtAl1998a,BrosensEtAl1998b}, 
Brosens, Devreese and Lemmens
explored this system and derive the canonical partition sum:
\begin{equation}
   Z_{N}\left(\beta, w\right) = \frac{\sinh^{D}\left( \beta\hbar w/2\right)}
   {\sinh^{D}\left( \beta\hbar\Omega/2\right)} \mathbb{Z}_{N}(w), \label{Z1}
\end{equation}
with
\begin{align}
   \mathbb{Z}_{N}(w) &  =\sum_{\substack{\{M_{1},M_{2},...,M_{N}\}\\
   {\textstyle\sum}
\ell M_{\ell}=N}} \left( 
{\prod_{\ell}}
   \frac{\xi^{M_{\ell}(\ell-1)}}{M_{\ell}!\ell^{M_{\ell}}}
   \frac{1}{\left[2\sinh\left(  \ell\beta\hbar w/2\right)  
   \right] ^{2M_{\ell}}}\right).
   \label{Z2}
\end{align}
Here $D$ is the 
number of spatial dimensions, and $\xi=-1$ for fermions and $\xi=+1$ for
bosons. The prefactor in front of $\mathbb{Z}_N$ in expression
(\ref{Z1}) takes care of the center of mass
degrees of freedom.
The summation runs over all sets of natural numbers $\{M_{1}%
,M_{2},...,M_{N}\}$ satisfying the condition that $%
{\textstyle\sum_{\ell=1}^{\infty}}
\ell M_{\ell}=N$. This restriction makes the sum impossible to evaluate
directly. However, it can be lifted by introducing the generating function%
\begin{equation}
\mathcal{G}_{0}(u)=\sum_{n=0}^{\infty}u^{n}\mathbb{Z}_{n}(w).\label{grand0}%
\end{equation}
The summation is taken while keeping $w$ fixed. It is clear that
$\mathbb{Z}_{n}(w)$ is the canonical partition function of $n$ non-interacting
identical particles in a harmonic confinement of frequency $w$. The generating
function evaluates to
\begin{equation}
\mathcal{G}_{0}(u)=\exp\left\{  \xi%
{\displaystyle\sum\limits_{\ell=1}^{\infty}}
\dfrac{\left(  \xi ub\right)  ^{\ell}}{\ell(1-b^{\ell})^{D}}\right\},
\label{grand}%
\end{equation}
where
\begin{equation}
b=\exp\left\{  -\beta\hbar w\right\}.
\end{equation}
In order to retrieve $\mathbb{Z}_{N}$ from $\mathcal{G}_{0}(u)$ one can use
the formula%
\begin{equation}
\mathbb{Z}_{N}=\frac{1}{N!}\left.  \frac{d^{N}\mathcal{G}_{0}}{du^{N}%
}\right\vert _{u=0}.
\end{equation}
This leads to a recursion formula as shown in \cite{BrosensEtAl1997a}%
\begin{equation}
\mathbb{Z}_{N}=\dfrac{1}{N}%
{\displaystyle\sum\limits_{m=0}^{N-1}}
\left(  \dfrac{b^{(N-m)/2}}{1-b^{N-m}}\right)  ^{D}\mathbb{Z}_{m}.
\end{equation}
In appendix~\ref{app:BDLA} we outline the algorithm we have implemented to generate the
list of $\mathbb{Z}_{N}$\thinspace's from this recursion formula in a
numerically stable way. We apply the BDL formalism for bosons ($\xi=1$) in
$D=2$ dimensions\cite{TempereDevreese1997}.

\bigskip

It would be tempting to identify $\mathcal{G}_{0}(u)$ defined above 
with a grand-canonical partition function $\Xi(u)$ (up to the correction
factor for the center of mass degree of freedom). As noted in \cite{BrosensEtAl1998a,LemmensEtAl2003},
this is wrong. In the sum for the generating function 
$\mathcal{G}_{0}(u)$ the frequency $w$ is kept fixed.
However, the interactions change the spectrum, so that systems with
a different number of particles have a different basic frequency 
$\hbar w$ for the harmonic oscillator. This is
typical for interacting systems: adding more particles changes the effective
single-particle levels. Indeed, mean-field shifts are an example of this. 
Usually, in defining the grand-canonical partition sum, this effect is 
ignored and one assumes that for large enough numbers of particles, 
adding or removing a few particles only leads to negligible changes 
in the effective single-particle spectrum.

For smaller numbers of atoms this is clearly not true and it is necessary to
go back to the original definition of the grand-canonical partition sum as an
ensemble sum. Consider an open subsystem where interactions are present, and a
reservoir (a large box of ideal gas) that can supply or remove particles from
the subsystem. The grand-canonical partition sum is then%
\begin{equation}
\Xi=\sum_{n}\sum_{E_{s}(n)}\exp\left\{  -\beta\left[  E_{s}(n)-\mu n\right]
\right\} \label{xi01}
\end{equation}
Here, the grand canonical ensemble of microstates of the entire system is
subdivided in canonical ensembles where the subsystem has $n$ particles. 
Each of these canonical ensembles can be further subdivided into
microcanonical ensembles according to the possible energy states $E_{s}$ of
the subsystem. However, the list of possible energy states does not need to be
the same for each canonical ensemble, so we have $E_{s}(n)$. In our
example, the values of $E_{s}$ are given by $\hbar w_{n}(\nu+1)$ with
$\nu\in\mathbb{N}$ and $w_{n}=\sqrt{\Omega^{2}+n\omega^{2}}$. The
grand canonical partition sum corresponding to (\ref{xi01}) can be written as
\begin{equation}
  \Xi(z)=\sum_{n=0}^{\infty}z^{n} Z_{n}(\beta,w_{n}).
  \label{xi}
\end{equation}
Given the list of $\mathbb{Z}_{n}$'s, and a fixed $N=\left\langle
n\right\rangle $, the fugacity $z$ is found by numerically solving
\begin{equation}
\left\langle n\right\rangle =\frac{1}{\Xi(z)}\sum_{n=0}^{\infty}%
nz^{n} Z_{n}(\beta,w_{n}).
\end{equation}
Note that, in general, not for every $N$ a solution exists for $z$ (for
example, in fermionic systems with partially filled degenerate energy 
levels\cite{LemmensEtAl2003}). However, even in these cases the generating
function formalism can still be used to calculate the exact canonical partition
sums.

Although $\Xi(z)$ is clearly different from $\mathcal{G}_{0}(u)$ defined
above, there is a link. When the average number of particles in the subsystem
$N=\left\langle n\right\rangle $ becomes large, and the temperature is
above the condensation temperature for bosons, the summation above is sharply
peaked around $N$. Only terms with $n\approx N$ contribute, so that often
one even approximates the sum by a single term, $\Xi(z)=z^{N}\mathbb{Z}%
_{N}(w_{N})$, the saddle-point approximation. This approximation yields the
well-known free energy formula $\Omega(\mu)=F(N)-\mu N$. 
This formula is only true at the
saddle-point level, as the neglected terms generate a correction term to it.
However, it means that if $N$ is large enough (and the temperature is above
the critical temperature of condensation), then in the small range of relevant
$n$'s in the summation one may keep $w_{n}\approx w_{N}$ fixed. Only
under these conditions $\Xi(z)\approx\mathcal{G}_{0}(z)$. When the conditions
are violated\ (for example when studying condensate fluctuations at low
temperatures), this is no longer true. This has a strong effect on the
fluctuations 
$\Delta N=( \langle n^{2}\rangle - \langle n \rangle ^{2} )^{1/2}$.

\bigskip

\section{\label{sec:results}Results}
\subsection{Overview of Sec.~\ref{sec:results}}
In Sec.~\ref{sub:total} we use the fact that for very low
temperatures, $k_{\rm B}T \ll \hbar \omega$, the model of
Ref.~\cite{IdziaszekEtAl1999} introduced in Sec.~\ref{sub:modelCW}
becomes exact to make analytical predictions for the size of the
fluctuations at low temperatures.

Temperatures are given in units of the condensation temperature of a non-interacting Bose gas in a two-dimensional harmonic oscillator (in the version which can be found, e.g., in Ref.~\cite{WeissWilkens1997} and references therein)
\begin{equation}
\label{eq:T0}
k_{\rm B}T_0 \equiv \hbar\omega{\frac {\sqrt {6}\sqrt {{\langle n\rangle}}}{\pi }}
\end{equation}

Both for the total number fluctuations (Sec.~\ref{sub:total}) and for condensate fluctuations (Sec.~\ref{sub:condensate}) as well as for the kurtosis (Sec.~\ref{sub:kurtosis}) the qualitative statements of the BDL model agree with the fundamentally different LTI model: on the one hand, for temperature $T\gtrapprox T_0$, a weakly interacting grand-canonical BEC essentially behaves like a non-interacting grand-canonical BEC. On the other hand, for low temperatures the deviations are very large --- the weakly interacting grand-canonical BEC behaves more like a canonical BEC.

\subsection{\label{sub:total}Total number fluctuations within the grand-canonical ensemble}


The LTI model allows to derive analytical results for the total number fluctuations for very low temperatures ($k_{\rm B}T \ll \hbar \omega$; $\exp(-\beta\hbar\omega)\ll 1$).
In this limit, Eq.~(\ref{eq:rec}) simplifies to
\begin{equation}
\Zce_N^{\rm(ex)}(\beta) \simeq 0,\quad N>0.
\end{equation}
Furthermore Eq.~(\ref{eq:Zcetotal}) becomes
\begin{equation}
\Zce_N(\beta) \simeq \exp\left[-\beta\alpha N(N-1)\right]
\end{equation}
and Eq.~(\ref{eq:Zgcetotal}) now reads
\begin{align}
\Zgce(\beta) &\simeq \sum_{n=1}^{\infty}z^n\exp\left[-\beta\alpha
  n(n-1)\right]\\
&\simeq \int_{-\infty}^{\infty}dn z^n\exp\left[-\beta\alpha
  n(n-1)\right]\label{eq:integral}\\
&=\frac{\sqrt {\pi }}{\sqrt{\beta\alpha}}\exp\left(\frac 14\frac {\left( \ln  \left( z
 \right) +{\beta\alpha} \right) ^{2}}{\beta\alpha}\right).
\end{align}
This analytical result can be used to obtain
\begin{align}
\langle n\rangle &= -z\frac{\partial}{\partial z}\ln[\Zgce(\beta) ]\\
&=  -1/2-{\frac {\ln  \left( z \right) }{{2\beta\alpha}}}
\end{align}
and, in particular
\begin{align}
\langle \Delta n^2\rangle &= z\frac{\partial}{\partial z} z\frac{\partial}{\partial z}\ln[\Zgce(\beta) ]\\
& =  \frac{k_{\rm B}T}{2\alpha}
\label{eq:DeltaNsqAna}
\end{align}
Thus, in the limit $k_{\rm B}T \lessapprox \alpha\ll \hbar \omega$, the
variance of the total number of particles is independent of $\langle
n\rangle$ and goes to zero. This implies that canonical and grand-canonical
ensemble become equivalent again.  Therefore, the we predict that the
grand-canonical  condensate
fluctuations vanish when the temperature goes to zero. 

The limit $\alpha \to 0$ has to be taken with care in
Eq.~(\ref{eq:DeltaNsqAna}): the fact that the relative fluctuations
$\Delta n/\langle n\rangle$ seem to diverge only signifies that the
approximation~(\ref{eq:integral}) breaks down.

However, as long as $k_{\rm B}T \ll \hbar \omega$, 
$\Delta n/\langle n\rangle<1$ and $T<T_0$, $\langle n\rangle (\langle n\rangle-1)\alpha<\hbar\omega$ the fluctuations of the
total number of bosons is given by Eq.~(\ref{eq:DeltaNsqAna}).

\begin{figure}
\includegraphics[width=\linewidth]{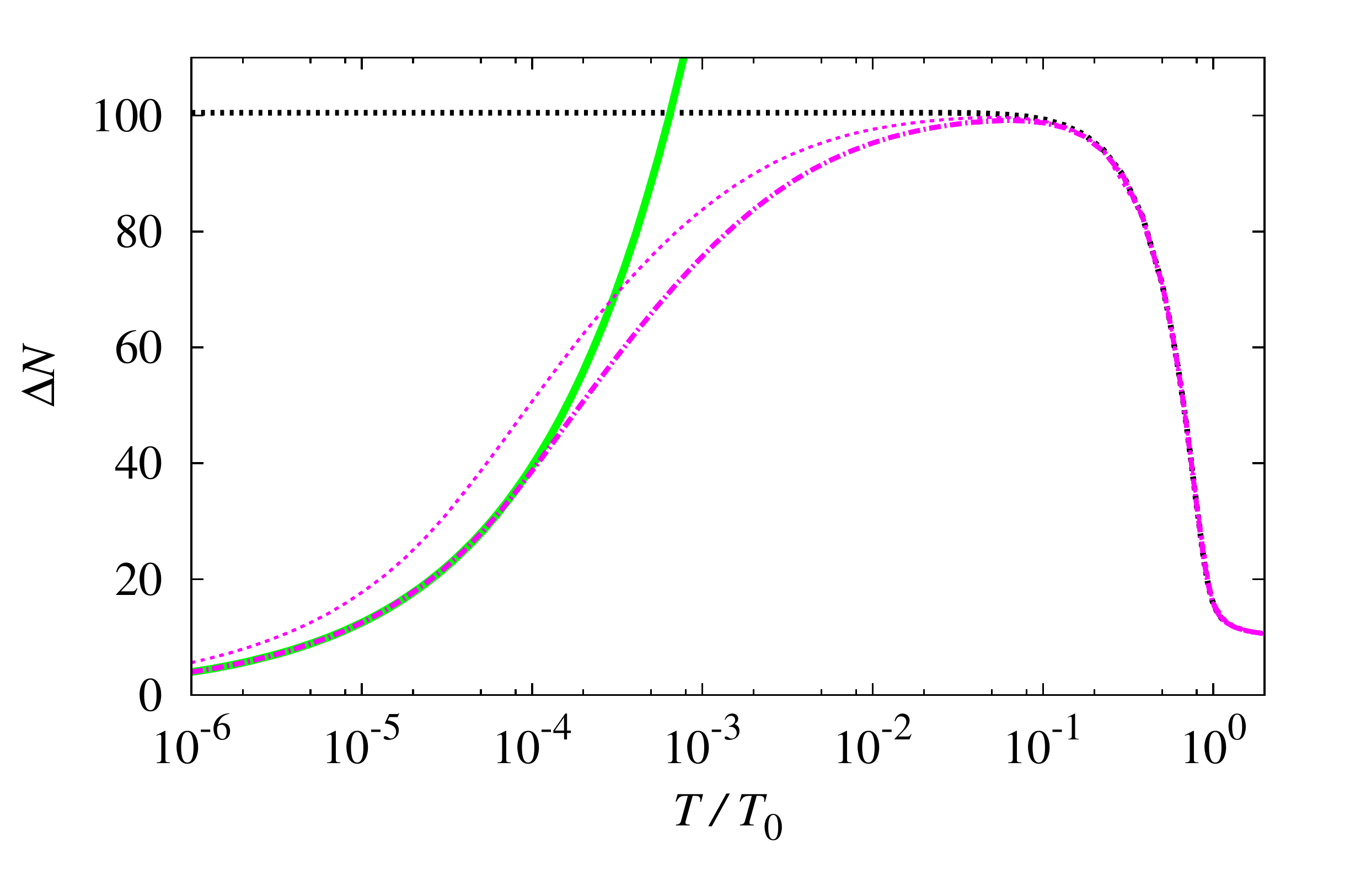} 
\caption[test]{\label{fig:LTI1}(Color online) The fluctuation $\Delta N$ in the total number of particles in the
subsystem within the LTI model. Thick magenta/dark gray dash-dotted curve: grand canonical ensemble $N_{max}=2000$, $\langle N \rangle = 100$, $\alpha = 1/4000000$. For very low temperatures, this agrees with the analytic prediction~(\ref{eq:DeltaNsqAna}) (green/light gray solid curve); for higher temperatures with the non-interacting grand-canonical ensemble (black dashed curve). Thin magenta/dark gray dashed curve: grand canonical ensemble for the same parameters as the thick dash-dotted curve except for a reduced $\alpha = 1/8000000$. For both interaction parameters predictions of all three versions of the LTI model are plotted; the data lie on top of each other.}
\end{figure}

While the analytical result of Eq.~(\ref{eq:DeltaNsqAna}) can only be valid for low temperatures, in this temperature regime it agrees well with the full numerical results within the LTI model shown in Fig.~\ref{fig:LTI1}. 
In order for the model to be consistent, we choose interactions that are weak in the sense that the sums in Eq.~(\ref{eq:ZGCE}) have converged at \mbox{$N\approx N_{\rm max}/2$} \textit{and}\/ $E_0(N_{\rm max}/2)\lessapprox \hbar \omega$.

To show the effect of interactions in the BDL model, we choose a 
grand canonical ensemble where
the average number of particles is fixed at $\left\langle n\right\rangle
=100.$ Taking $\omega=0.01$ for the interaction strength (and $\Omega=1$ as
energy unit), we use the number-dependent frequency $w_{n}=\sqrt{\Omega
^{2}+n\omega^{2}}$ to calculate the expectation value 
$\left\langle n^{2}\right\rangle $, 
and the corresponding $\Delta N$. For the corresponding
non-interacting case, we use the number-independent value $w_{100}$ for the
frequency at any $n$. If the sums are very sharply peaked around $n=100$, then
the effect of the interactions on the fluctuations will hence be small.
However, if the sums broaden out and many terms need to be taken into account
to get the total, then there will be a significant difference. So, we expect
that for higher temperatures, both results will coincide, whereas for
temperatures below the condensation temperature, differences may occur.
\begin{figure}[ptb]%
\centering
\includegraphics[width=\linewidth
]{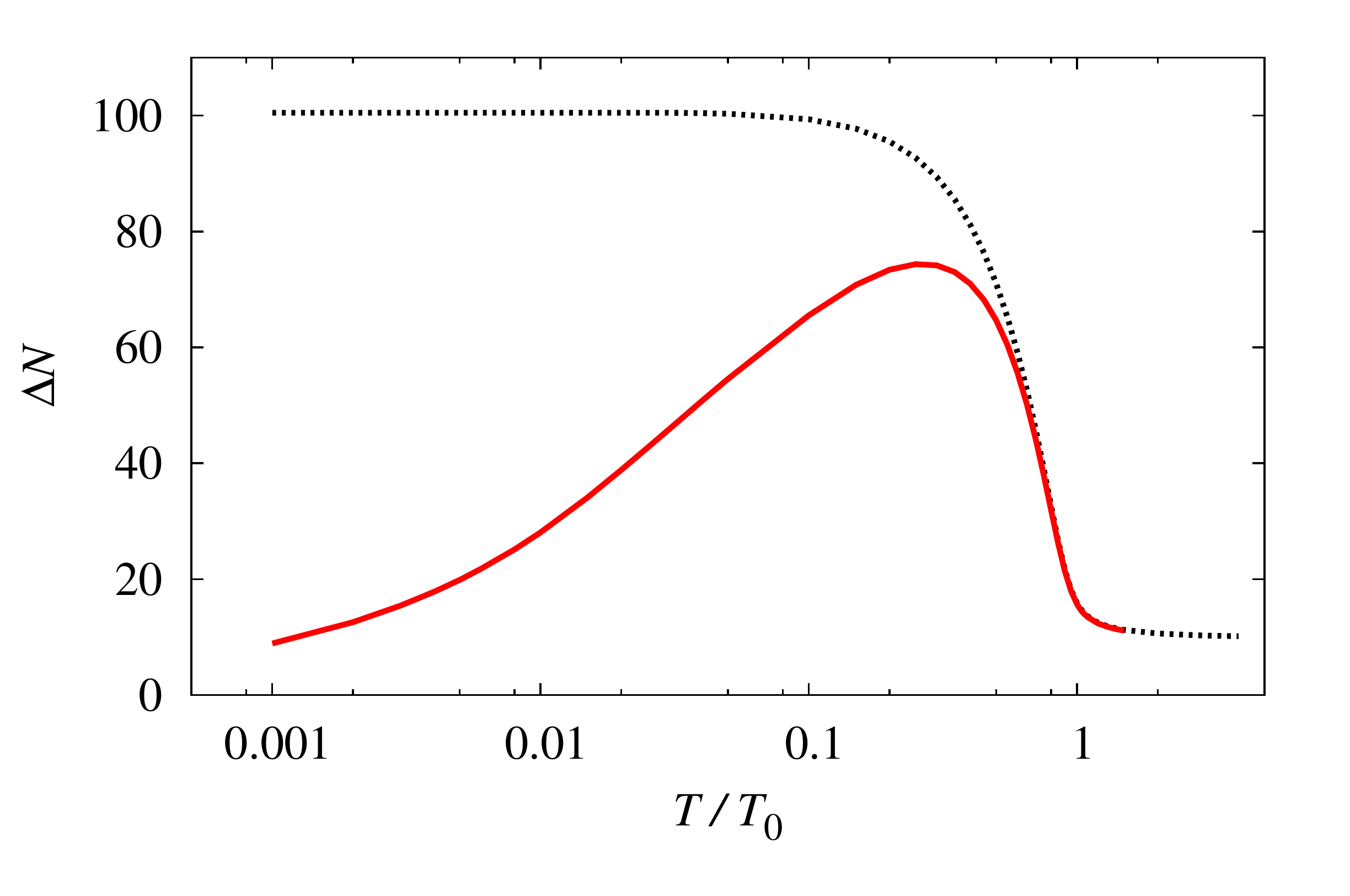}
\caption{(Color online) The fluctuation $\Delta N$ in the total number of particles in the
subsystem within the BDL model, in the grand-canonical ensemble is plotted here as a function of
temperature. The chemical potential is set such that the average number of
particles in the subsystem is $\left\langle n\right\rangle =100$. The
temperature unit is $T_{0}=\hbar w_{100}\sqrt{6\times100}/\pi$, where
$w_{100}=\sqrt{\Omega^{2}+100\omega^{2}}$ and $\Omega=1,\omega=0.01$. The
dashed black curve keeps $w=w_{100}$ constant, the full red (black)
curve takes into account interactions and uses $w_{n}$. }%
\label{figgctotal}%
\end{figure}
The results are shown in figure \ref{figgctotal}. For temperature unit we use
$T_{0}=\hbar w_{100}\sqrt{6\times100}/(\pi k_{B})=7.797$ $\hbar w_{100}/k_{B}%
$. At large temperatures, the fluctuations tend to $\Delta N=\sqrt{N}$ as one
expects in the classical limit (this is indicated by a horizontal line at
$\Delta N=10$ in the figure). When no interactions are present (dashed black
curve), the grand-canonical fluctuations on the total number tend to $\Delta
N=\sqrt{N\left(  N+1\right)  }$ as the temperature goes to zero (there is a
horizontal line at $\Delta N=100$ in the figure). When interactions are
present, this reduces the grand-canonical fluctuations for temperatures below
$T_{0}$. Nevertheless, there is an intermediate regime where the fluctuations
are still substantial, reaching $\Delta N\approx70$ out of $N=100$ particles.
In the absence of strong correlations, the fluctuation in the number of
particles in any given energy level should be smaller than the fluctuation in
the total number of particles. In particular, we expect that the fluctuation
in the condensate number is bound by $\Delta N$. Hence, interactions 
also prevent grand-canonical condensate fluctuations from becoming as large 
as the total number of particles.

\bigskip

\subsection{\label{sub:condensate}Condensate number fluctuations}


For our investigation of the condensate number fluctuations, we start with numerical results obtained within the BDL model.
Using again $\left\langle n\right\rangle =100$ to fix $z$, we can now
calculate the grand-canonical expectation value for the number of condensate
atoms as%
\begin{equation}
\left\langle n_{0}\right\rangle ^{\text{gc}}=\frac{1}{\Xi(z)}%
{\displaystyle\sum\limits_{n=0}^{\infty}}
\left\langle n_{0}\right\rangle _{n}^{{\text{can}}}z^{n}Z_{n}(\beta,w_{n}).
\end{equation}
Here, $\left\langle n_0\right\rangle _{n}^{{\text{can}}}$ is the expectation
value of the number of condensate atoms in the \emph{canonical} ensemble with
$n$ particles, and $\Xi(z)$ is given by expression (\ref{xi}).
Within the BDL
model, the canonical condensate number can be expressed as a recursion
relation\cite{BrosensEtAl1997a}, as outlined in appendix~\ref{app:BDLB}. 
The algorithm for computing the recursion
relation is outlined in appendix~\ref{app:BDLA}. Similarly, $\left\langle n_{0}%
^{2}\right\rangle _{n}^{{\text{can}}}$ can be computed, and used to find
$\left\langle n_{0}^{2}\right\rangle ^{\text{gc}}$, allowing us to find the
grand-canonical fluctuations for the number of condensate atoms,%
\begin{equation}
\left(  \Delta n_{0}\right)  ^{\text{gc}}=\sqrt{\left\langle n_{0}%
^{2}\right\rangle ^{\text{gc}}-\left(  \left\langle n_{0}\right\rangle
^{\text{gc}}\right)  ^{2}}.
\end{equation}
As before, to switch off the effects of interaction, we work at a fixed
frequency $w_{100}$ rather than an $n$-dependent frequency.%

\begin{figure}[ptb]
    \includegraphics[width=\linewidth]{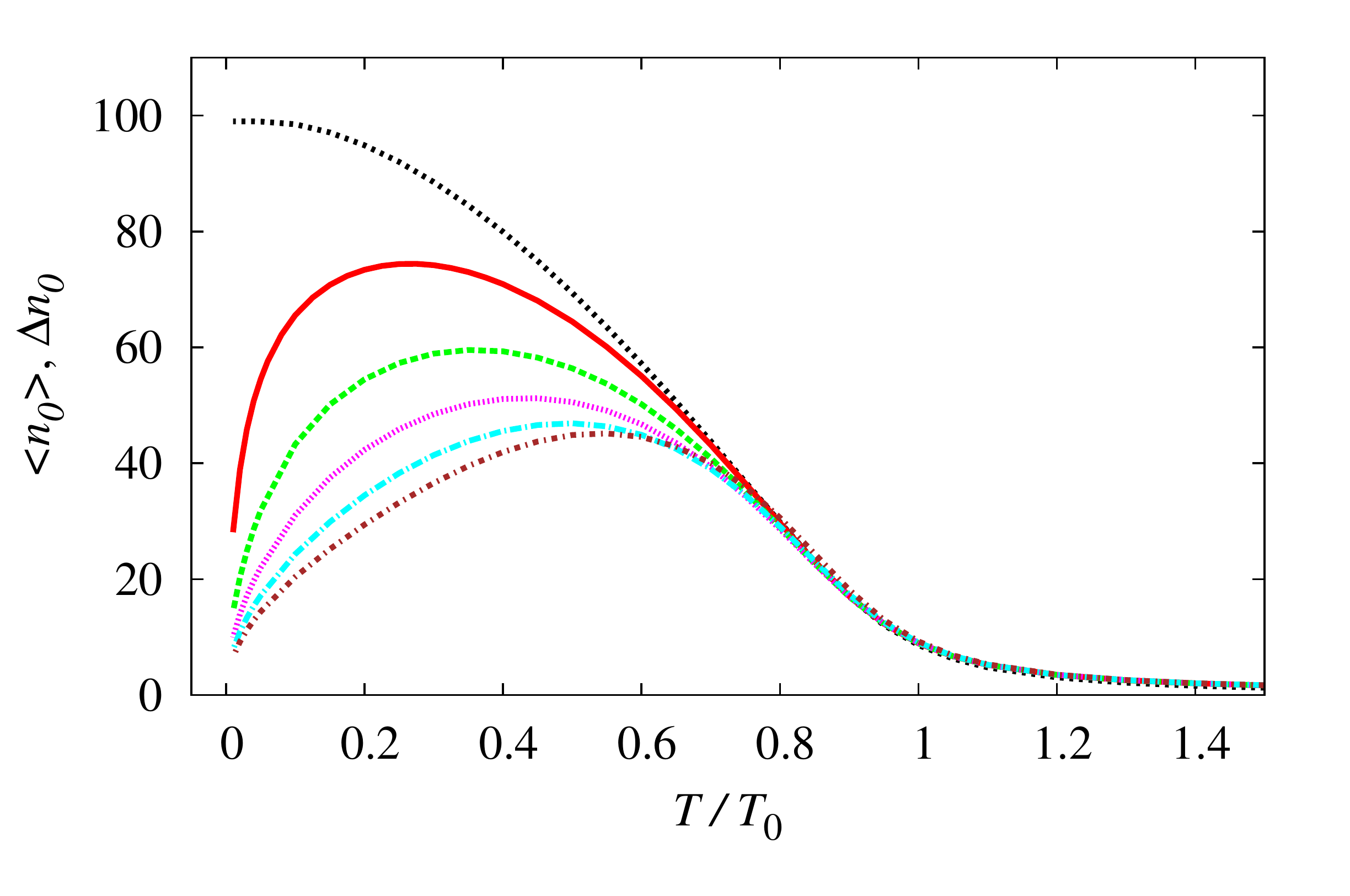}
    \caption{The statement that the stronger the interaction the smaller are interactions is also true in the BDL model. Following the curves from the left-hand side of the figure yields: the top curve shows the condensate fraction, all other curves show the condensate fluctuations for (from top to bottom): $\omega=0.01,0.02,0.03,0.04$ and $0.05$.}
    \label{fig:BDLvariousInteractions}
\end{figure}

\begin{figure}[h!]%
\centering
\includegraphics[width=\linewidth
]{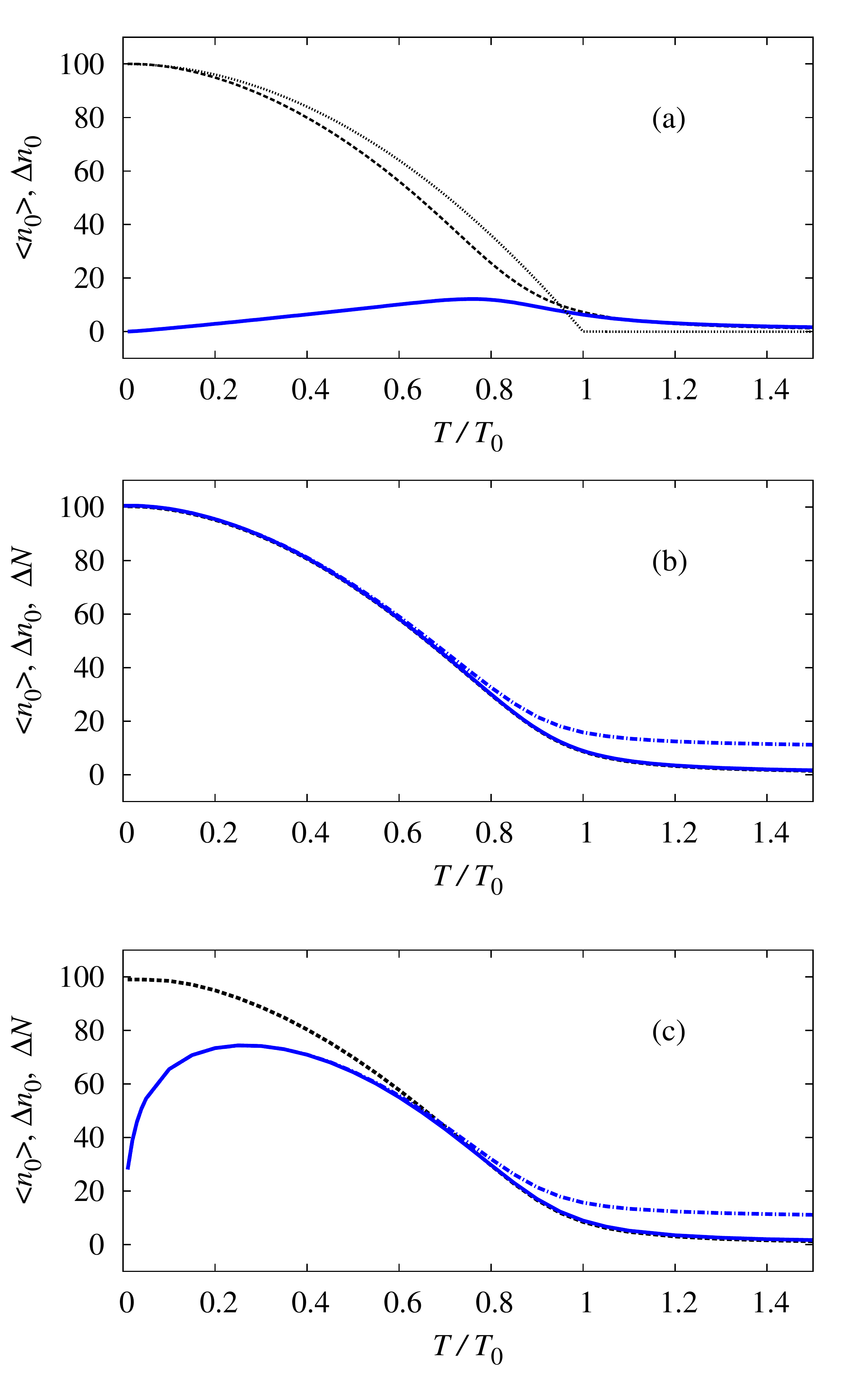} 
\caption[test]{(Color online) In each panel results within the BDL model are shown, namely the number of condensate atoms $n_{0}$ (dashed black
line), and the fluctuation $\Delta n_{0}$ of this condensate number (full blue
line), are plotted as a function of temperature. The top panel shows the
result in the canonical ensemble with $100$ particles. The middle panel shows
the results in the grand-canonical ensemble, for the case without interactions
-- the fluctuations is as large as the condensate number. The bottom panel
illustrates the effect of interactions ($\omega=0.01$ in the model) on the
grand-canonical fluctuations. The dotted black line in panel (a) show the result for
$n_{0}$ in the thermodynamic limit. The dash-dotted blue (black) line in the bottom
two panels shows the fluctuations for the total number of particles in the
grand-canonical ensemble.}
\label{figgccond}%
\end{figure}

The results are shown in Figs.~\ref{fig:BDLvariousInteractions} and \ref{figgccond}. Figure~\ref{fig:BDLvariousInteractions} shows that the stronger the interactions the lower the fluctuations is also true in the BDL model.
In Fig.~\ref{figgccond} three cases are compared:
(a) the canonical ensemble for $100$ particles, 
(b) the grand-canonical ensemble without interactions (with 
($\left\langle n\right\rangle=100$), and  
(c) the grand-canonical ensemble with interactions ($\omega=0.01$ and again 
$\left\langle n\right\rangle=100$). The dashed line shows the result for the condensate
number, closely equal in all three cases, and showing some finite-number effects when
compared to the result in the thermodynamic limit (dotted line in panel (a)). 
The full line shows
the condensate fluctuations as a function of temperature, and for this quantity it is
clear that the results strongly differ. In the canonical ensemble, fluctuations remain
small, and are only weakly enhanced close to the critical temperature. Without
interactions, in panel (b), it is apparent that the condensate
fluctuations are as large as the condensate number. Interactions suppress the
grand-canonical fluctuations at low enough temperature: in panel (c) the fluctuations
decrease when the temperature dips below $0.3 T_0$. Numerical inaccuracy prevents to
compute the $T=0$ results, so the lowest temperature shown is $T=0.01$. The dash-dotted
curve in panels (b) and (c) shows the grand canonical fluctuation for the total number
of particles. Below the condensation temperature, the fluctuation in total number is
dominated by the contribution of the condensate fluctuations. Near the critical
temperature and above it, the condensate fluctuations dip below the fluctuation in the
total number of atoms, as expected. At high temperatures, the standard square-root
fluctuations are retrieved. From panel (c) it is also clear that interactions 
suppress both fluctuations in the total number of particles, but also condensate fluctuations.

\bigskip

\begin{figure}
\includegraphics[width=\linewidth]{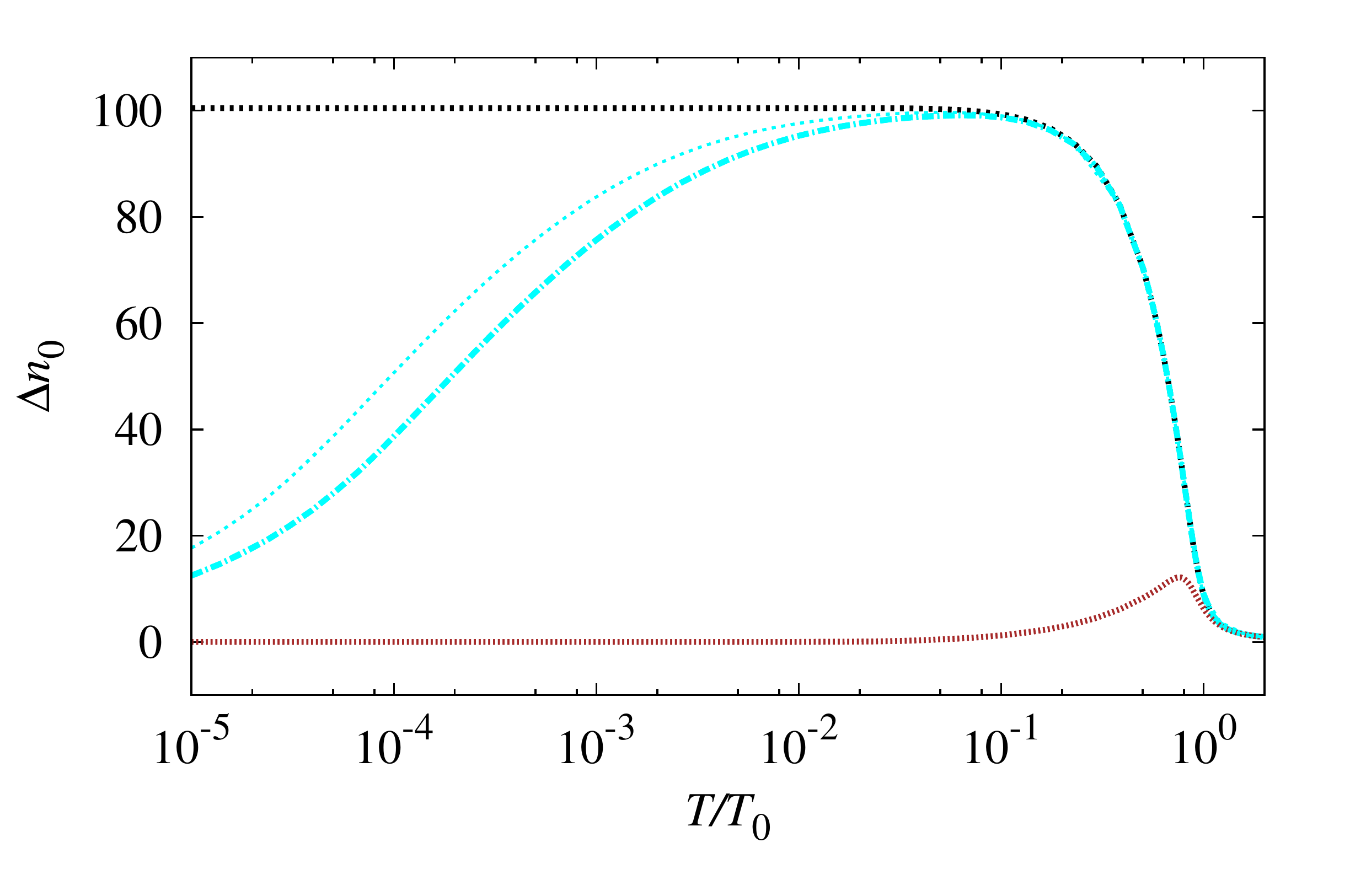}
\caption[test]{\label{fig:LTI2}(Color online) The condensate number fluctuation $\Delta n_0$ within the LTI model. Thick light blue/light gray dash-dotted curve: grand canonical ensemble $N_{max}=2000$, $\langle N \rangle = 100$, $\alpha = 1/4000000$. For higher temperatures, this again agrees with the non-interacting grand-canonical predictions (black dashed curve). As for the fluctuations of the total number, for lower temperature the system behaves more like a canonical ensemble (black/brown dotted curves) for which interacting and non-interacting curves lie on top of each other.   Thin light blue / light gray dashed curve: decreasing the interaction in the grand canonical enesmble by a factor of two again shifts the curve toward higher fluctuations (cf.\ Fig.~\ref{fig:LTI1}). As in Fig.~\ref{fig:LTI1}, the predictions of all three versions or the LTI model lie on top of each other.}
\end{figure}

The LTI model again yields results that are qualitatively very similar to the BDL model.
Figure~\ref{fig:LTI2} displays the condensate fluctuations within the LTI model. Although this model considerably differs from  the BDL model, the qualitative statements about the behavior at low and high temperatures are the same as in the BDL model (Fig.~\ref{figgccond}): for low temperatures, interaction dramatically changes the condensate fluctuations to a behavior close to the canonical ensemble: the condensate fluctuations vanish at low temperatures (as was to be expected from the vanishing fluctuations in the total number of photons, cf.~Figs.~\ref{fig:LTI1} and 
\ref{figgctotal}).

\bigskip

\subsection{\label{sub:kurtosis}Kurtosis and non-Gaussian fluctuations of the total number
of particles}


\begin{figure}[ptb]%
\centering
\includegraphics[width=\linewidth
]{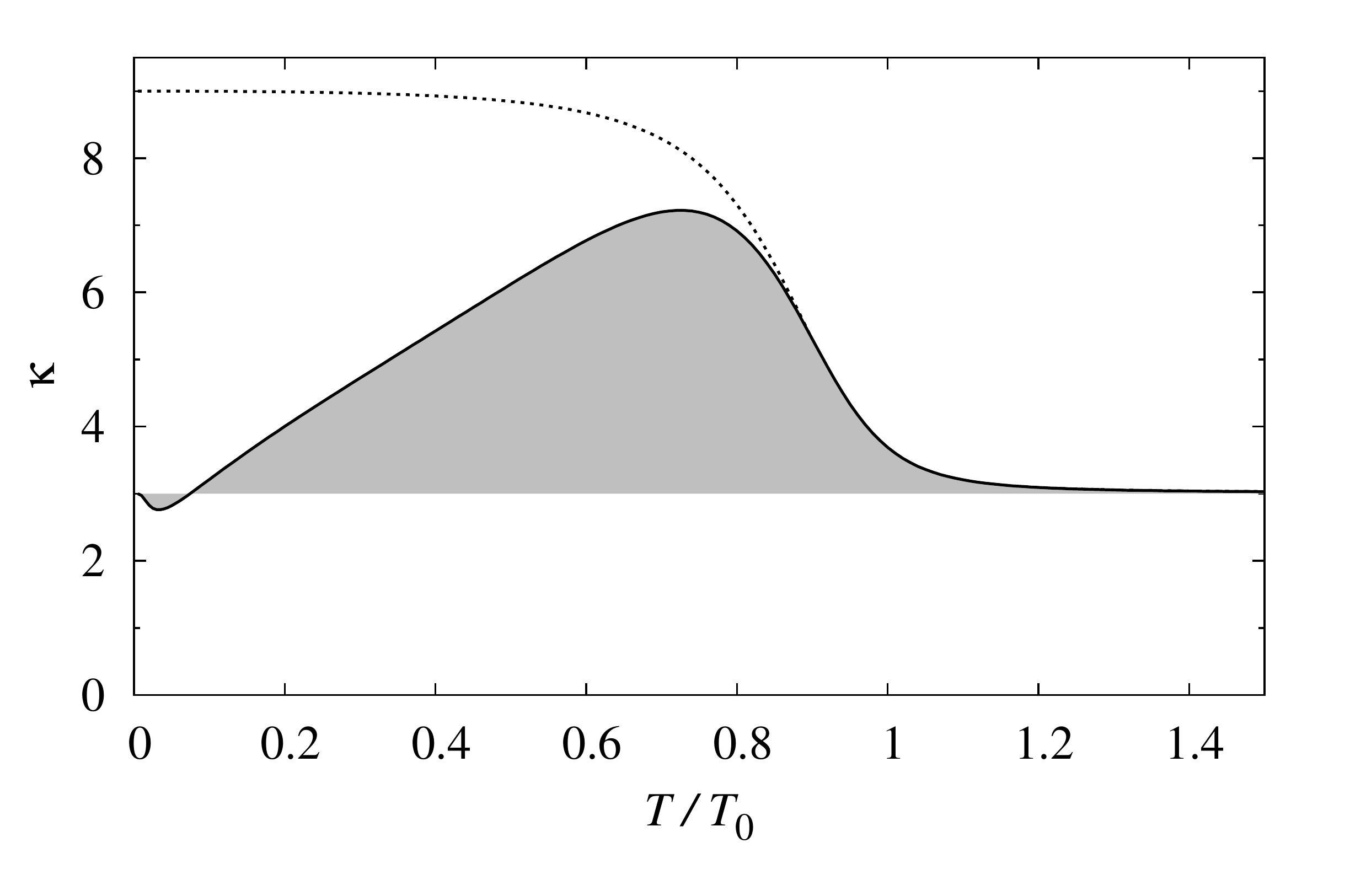}%
\caption{Kurtosis~(\ref{eq:kurtosis}) of the total number fluctuation distribution in the grand
canonical ensemble with the BDL model, as a function of temperature, and for $\left\langle
n\right\rangle =100$. The full line is for the case with interactions
($\omega=0.01$), whereas the dashed line is for the ideal gas. The value $\kappa=3$ would correspond to Gaussian fluctuations; it is highligthed via the shaded region.}%
\label{figkurt}%
\end{figure}
The \textquotedblleft classical\textquotedblright\ fluctuations of the total
number of atoms in an open subsystem are Gaussian, with a width equal to
$\sqrt{N}$. When the temperature of a Bose gas
drops below the critical temperature, these fluctuations increase strongly,
and the standard deviation becomes as large as the total number of particles,
$\sqrt{N \left( N+1\right)}$. Does this correspond to a wider Gaussian, 
or do the fluctuations obtain a non-Gaussian character? A widely used
(see, e.g.,
Refs.~\cite{WeissHolthaus2002,BeckCohen2003,BeckEtAl2005,BeckEtAl2006}
and references therein) quantifier of non-gaussianity is the excess kurtosis, $\gamma=\kappa-3$ with
the kurtosis itself defined by%
\begin{equation}
\label{eq:kurtosis}
\kappa=\frac{\left\langle \left(  n-N \right)
^{4}\right\rangle }{\left(  \Delta N\right)  ^{4}},
\end{equation}
i.e. the ratio of the fourth central moment to the second central moment
squared. Here again $N=\left\langle n\right\rangle$ and $\left( \Delta N\right)  ^{2}=\left\langle \left( n-N \right) ^{2}\right\rangle $. For a Gaussian
distribution, $\kappa=3$. There exist other methods to measure non-Gaussianity
(for example the entropy) that are less sensitive to outliers, but the
kurtosis is very easy to calculate in the present treatment, since we can 
compute any moment $\left\langle n^{p}\right\rangle ^{\text{gc}}$ of the
grand-canonical number distribution (as outlined in appendix~\ref{app:BDLB}). The excess kurtosis can be determined by repeated measurement of the number and building up statistics to estimate second and fourth moment. Applications in other fields, for example detecting "fat tails" in the velocity distributions of turbulent fluids \cite{BeckCohen2003} or in time series of financial instruments~\cite{MantegnaStanley2000}, show that using the fourth moment to estimate the non-Gaussian nature of the distribution is robust to other sources of noise, and more accessible than for example Castaign's measure~\cite{CastaingEtAl1990}.

Figure \ref{figkurt} shows the result for $\kappa$ as a function of
temperature. For the thermal Bose gas, at temperatures above $T_0$, the
fluctuations are indeed Gaussian and $\kappa=3$. Excess kurtosis appears below
$T_{c}$, indicating a more \textquotedblleft peaked\textquotedblright%
\ distribution with a \textquotedblleft fatter tail\textquotedblright. When no
interactions are present, this excess kurtosis remains present as the
temperature is lowered to zero. Including interactions ($\omega=0.01$ as
before) is seen to reduce the kurtosis back down to 3 as the temperature goes
to zero. So, interactions not only suppress the size of the fluctuations, they
also tend to keep the fluctuations Gaussian. This effect is more pronounced at
temperatures well below $T_0$. When the temperature is near to $T_0$ the
effect of the interactions is less. Note that there is a small leptokurtic
region at temperatures just above zero -- it is not clear whether this is an
artifact of the calculations (we take $n$ up to 2000 for the sum in the
grand-canonical ensemble) or whether this is real.

\begin{figure}
\includegraphics[width=\linewidth
]{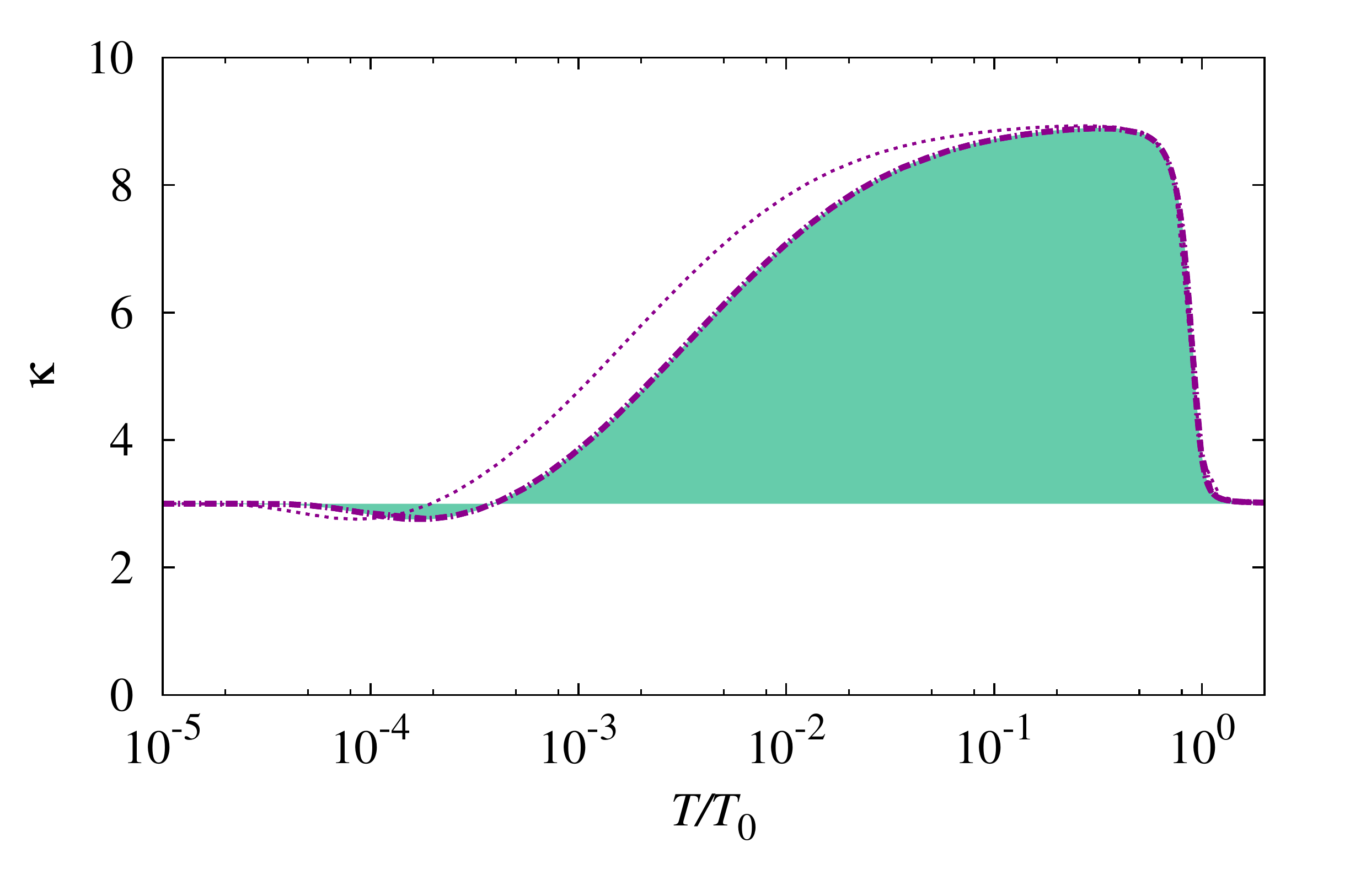}%
\caption{(Color online) Thick dash-dotted purple / black line:  kurtosis~(\ref{eq:kurtosis}) of the total number fluctuation distribution in the grand canonical ensemble with the LTI model. The shaded green / black area indicates if the kurtosis lies above or below the result for Gaussian fluctuations. Interaction parameters as in Fig.~\ref{fig:LTI1}; the thin dashed purple / black line again represents the parameters for which the interaction has been decreased by a factor of 2; for very low temperatures this decreases the kurtosis, for larger temperatures it increases the kurtosis. Qualitatively, the LTI model again predicts a behavior similar to the BDL model (Fig.~\ref{figkurt}). As in Figs.~\ref{fig:LTI1}  and \ref{fig:LTI2}, the predictions of all three versions or the LTI model lie on top of each other.}
\label{fig:LTI3}
\end{figure}

The LTI model (Fig.~\ref{fig:LTI3}) again predicts a qualitatively very similar behavior to the BDL model (Fig.~\ref{figkurt}).

\bigskip

\section{\label{sec:conclusions}Conclusion}


Motivated by the landmark experiment observing grand-canonical statistics in photon condensates~\cite{SchmittEtAl2014}, we revisit differences between statistical ensembles when describing Bose-Einstein condensates. The focus if the current paper lies on microscopic modelling of thermodynamic properties for small but experimentally accessible particle numbers. For an ideal Bose gas the grand-canonical ensemble predicts rms fluctuations for both the total number of particles and for the number of particles in the condensate that are larger than the average number of particles (see Figs.~\ref{fig:LTI1}, \ref{figgctotal} and \ref{fig:LTI2}). While this clearly is unphysical for an atomic Bose-Einstein condensate with constant atom numbers, photon condensates open an entirely new world by making highly fluctuating total photon numbers possible.

However, the realization of a grand-canonical photon condensate does not automatically imply that the textbook treatment of an ideal gas is valid. Thus, we model \textit{weakly interacting}\/ photon condensates in order to investigate both condensate fluctuations and fluctuations of the total number of particles.
The exactly solvable model systems of
Brosens, Devreese and Lemmens (BDL model)~\cite{BrosensEtAl1997a,BrosensEtAl1997b} 
and a Low Temperature Interaction model (LTI model) based on~\cite{IdziaszekEtAl1999}. The latter model becomes exact in the limit of low temperatures for pairwise interactions. As our two models are fundamentally differerent from each other, this helps to identify model-independent properties of, for example,  fluctuations of the total number of particles.

Our paper focuses on a grand-canonical Bose gas interactions that are so weak that the canonical fluctuations (for fixed total particle numbers) match well the
results for a non-interacting system of bosons. The main results are listed below:

\begin{enumerate}
\item In the recent experiment of Ref.~\cite{SchmittEtAl2014} it is
mentioned that the statistics can be influenced by changing the
effective size of the reservoir, by changing the number of dye
molecules. However, as in the variational treatment of 
Ref.~\cite{vanderWurffEtAl2014}, our microscopic
treatment finds that already \textit{\textbf{weak interaction changes the fluctuations
dramatically for low temperatures:}}\/ rather than $\langle\Delta
N_0^2\rangle\sim \langle N_0\rangle^2$ as predicted for the ideal gas, 
the grand-canonical fluctuations vanish at low temperatures.

    \item  The \textit{\textbf{interactions come into play essentially when we move to the
grand-canonical ensemble}}. Thus, the textbook example for treating Bose-Einstein condensation, the ideal Bose gas, is a too idealized example to correctly describe the grand-canonical fluctuations investigated experimentally in Ref.~\cite{SchmittEtAl2014}. Slightly below the condensation temperature we still find very large rms fluctuations $\propto N$. As our paper focuses on experimentally realistic small particle numbers, this is very distinct from considerations arguing against the existence of anomalous fluctuations based on stability arguments in the thermodynamic limit~\cite{Yukalov2005,Yukalov2005b}. However, in agreement with~\cite{Yukalov2005,Yukalov2005b}, even very weak interactions considerably suppress fluctuations.

\item We find
that the \textit{\textbf{fluctuations are strongly non-Gaussian, showing substantial excess 
kurtosis}}. For low temperatures, even very weak interactions suppress both fluctuations and the non-Gaussian character of the fluctuations.
\end{enumerate}


For the BDL model, 
the fact that the center of
mass degree of freedom may have a different frequency from all the other
degrees of freedom does not alter the number of condensate particles, nor the
condensate fluctuation. When no interactions are present, it is known that the
condensate fluctuations in the grand-canonical ensemble become as large as the
total number of particles.

In the BDL model considered here, the role of
interactions is to change the spectrum of possible energy states of the
subsystem as a function of the number of particles in the subsystem. There is
a secondary effect of interactions, clear in the transition from
$\mathbb{Z}_{N}$ to $Z_{N}$, and which consists in endowing one degree of
freedom with frequency $\Omega$ rather than $w_{N}$. 
However, this single degree of freedom has only a negligible
effect on the results, even for $\left\langle n\right\rangle $ as low as $100$
-- whether we work with $\mathbb{Z}_{N}$ or $Z_{N}$, we find qualitatively the
same results. These results for the harmonically interacting model system show
that interactions reduce the number fluctuations in the grand-canonical ensemble, 
and make the fluctuations more Gaussian.

The data presented in this paper
 will be available soon
from~\url{https://collections.durham.ac.uk/files/vt150j289}
and from~\url{http://dx.doi.org/10.15128/vt150j289}~\cite{WeissTempere2016Data}.

\acknowledgments
 Insightful  
discussions with T.P.~Billam, F.~Brosens, J.T.~Devreese, S.A.~Gardiner,  L.~Lemmens,  W.~Magnus,  M.\ Weitz and M.~Wouters are gratefully acknowledged. 
CW thanks the
UK Engineering and Physical Sciences Research Council for funding (Grant No.~EP/L010844/, C.W.). 
JT acknowledges support from the Research
Foundation Flanders, project nrs. G.0429.15N, G.0119.12N, G.0122.12N,
and support from the Research Council of Antwerpen University.

\begin{appendix}

\section{\label{app:BDLA}Efficient recursion algorithms for BDL }

Retrieving $\mathbb{Z}(N)$ from $\mathcal{G}_{0}(u)$ leads to the following
recursion formula, as derived in \cite{BrosensEtAl1997a}:\
\begin{equation}
\mathbb{Z}_{N}=\dfrac{1}{N}%
{\displaystyle\sum\limits_{m=0}^{N-1}}
\xi^{N-m-1}\left(  \dfrac{b^{(N-m)/2}}{1-b^{N-m}}\right)  ^{D}\mathbb{Z}%
_{m}.\label{recurr}%
\end{equation}
with $\xi=1$ for bosons and $\xi=-1$ for fermions, in $D$ dimensions, and with
$b=e^{-\beta\hbar w}$. Note that $\mathbb{Z}_{0}=1$ and
\begin{equation}
\mathbb{Z}_{1}=\left[  2\sinh\left(  \beta\hbar w/2\right)  \right]  ^{-D}.
\end{equation}
Straightforwardly implementing the recursion formula leads to large numerical
inaccuracies. For bosons, there is a numerically fast and stable algorithm for
evaluating it, as communicated to one of us (JT) by F. Brosens. We are not aware of it
being described in detail in other publications, although it is mentioned in
\cite{BrosensEtAl1997b} where it is used to calculate the density and pair correlation in
the 3D Bose gas. To facilitate reproducing our results, we include the
description of the algorithm in this appendix. 

The recurrence relation is recast in a numerically more stable form by
introducing new quantities $\mathfrak{z}(j)$ (for $j=1,...,N$) defined by%
\begin{equation}
\mathbb{Z}_{N}=%
{\displaystyle\prod\limits_{j=1}^{N}}
\mathfrak{z}(j)\left(  \frac{b^{1/2}}{1-b^{j}}\right) ^{D}, \label{product}%
\end{equation}
where by definition $\mathfrak{z}(1)=1$. Introducing the above representation
of $\mathbb{Z}$ in the recurrence relation (\ref{recurr}) and isolating
$\mathfrak{z}(N)$ yields%
\begin{equation}
\mathfrak{z}(N)=\frac{1}{N}\sum_{m=0}^{N-1}\frac{(1-b^{N})^{D}}
{(1-b^{N-m})^{D}}%
{\displaystyle\prod\limits_{j=m+1}^{N-1}}
\frac{\left(  1-b^{j}\right)  ^{D}}{\mathfrak{z}(j)}. \label{zetarecur}%
\end{equation}
Once the list of $\mathfrak{z}(j)$'s is found, we can find
\begin{equation}
\ln\left(  \mathbb{Z}_{N}\right)  =\sum_{j=1}^{N}\left[  D\ln\left(
\frac{b^{1/2}}{1-b^{j}}\right)  +\ln\left[  \mathfrak{z}(j)\right]  \right],
\end{equation}
which in turn leads to a recursion relation for the $\ln(\mathbb{Z)}$'s$:$%
\begin{equation}
\ln\left(  \mathbb{Z}_{N}\right)  =D\ln\left(  \frac{b^{1/2}}{1-b^{N}}\right)
+\ln\left[  \mathfrak{z}(N)\right]  +\ln\left[  \mathbb{Z}_{N-1}\right]
,\label{zlist}%
\end{equation}
where we start with $\ln\left[  \mathbb{Z}_{0}\right]  =0$, and since
$\mathfrak{z}(1)=1$, $\ln\left[  \mathbb{Z}_{1}\right]  =D\ln\left[
b^{1/2}/(1-b)\right]  $. Note that this recursive formula will yield the list
$\mathbb{Z}_{m}(w)$ for a fixed $w$. Expression (\ref{zetarecur}) is
implemented through succesive multiplications and additions, by the following
algorithm which calculates $\mathfrak{z}(N)$ given $\mathfrak{z}%
(1)$,...,$\mathfrak{z}(N-1)$:%
\begin{equation}
\fbox{%
\begin{tabular}
[c]{l}%
$\text{start with }\left(  \frac{1-b}{1-b^{N}}\right)  ^{D},$\\
$\text{multiply by }\frac{\left(  1-b\right)  ^{D}}{\mathfrak{z}(1)}\text{ and
add }\left(  \frac{1-b}{1-b^{N-1}}\right)  ^{D},$\\
$\text{multiply by }\frac{\left(  1-b^{2}\right)  ^{D}}{\mathfrak{z}(2)}\text{
and add }\left(  \frac{1-b}{1-b^{N-2}}\right)  ^{D},$\\
...\\
$\text{multiply by }\frac{\left(  1-b^{N-2}\right)  ^{D}}{\mathfrak{z}%
(N-2)}\text{ and add }\left(  \frac{1-b}{1-b^{2}}\right)  ^{D}$\\
$\text{multiply by }\frac{\left(  1-b^{N-1}\right)  ^{D}}{\mathfrak{z}%
(N-1)}\text{ and add }1\text{,}$\\
$\text{divide by }N\left(  \frac{1-b}{1-b^{N}}\right)  ^{D}.$%
\end{tabular}
}\label{algozeta}%
\end{equation}
Note that for every $N$ we need to perform the entire algorithm: we have not
found a way to obtain $\mathfrak{z}(N)$ with less than $\mathcal{O}(N)$
additions and multiplications. The algorithm works for $N>1$ and should be
started with $\mathfrak{z}(1)=1.$

\bigskip

A similar algorithm can be found for canonical expectation values that can be
written as%
\begin{equation}
\mathbb{E}\left[  f\right]  =\sum_{m=0}^{N}f(m)\frac{\mathbb{Z}_m}%
{\mathbb{Z}_N}.%
\end{equation}
Both the expression for $\left\langle n_{0}\right\rangle _{N}^{{\text{can}}}$
and that for $\left\langle n_{0}^{2}\right\rangle _{N}^{{\text{can}}}$ are of
this form (see appendix~\ref{app:BDLB}). 
Then we rewrite the above using the product form (\ref{product}):%
\begin{equation}
\frac{\mathbb{Z}_m}{\mathbb{Z}_N}=%
{\displaystyle\prod\limits_{j=m+1}^{N}}
\frac{1}{\mathfrak{z}(j)}\left(  \frac{1-b^{j}}{b^{1/2}}\right) ^{D}.%
\end{equation}
Thus, we have
\begin{equation}
\mathbb{E}\left[  f\right]  =f(N)+\sum_{m=1}^{N-1}f(m)%
{\displaystyle\prod\limits_{j=m+1}^{N}}
\frac{1}{\mathfrak{z}(j)}\left(  \frac{1-b^{j}}{b^{1/2}}\right)  ^{D}%
+\frac{f(0)}{\mathbb{Z}_N}.%
\end{equation}
This can again be written as a series of summations and multiplications:%
\begin{equation}
\fbox{%
\begin{tabular}
[c]{l}%
$\text{start with }f(0),$\\
$\text{multiply with }\frac{1}{\mathfrak{z}(1)}\left(  \frac{1-b}{b^{1/2}%
}\right)  ^{D}\text{ and add }f(1),$\\
$\text{multiply with }\frac{1}{\mathfrak{z}(2)}\left(  \frac{1-b^{2}}{b^{1/2}%
}\right)  ^{D}\text{ and add }f(2),$\\
$\text{...}$\\
$\text{multiply with }\frac{1}{\mathfrak{z}(N)}\left(  \frac{1-b^{N}}{b^{1/2}%
}\right)  ^{D}\text{ and add }f(N).$%
\end{tabular}
}%
\end{equation}
This avoids the calculation of the canonical partition sums altogether. All we
need is the list of $\mathfrak{z}(j)$'s obtained from (\ref{zetarecur}) with the
algorithm (\ref{algozeta}).

\bigskip

\section{\label{app:BDLB}canonical condensate fluctuations in BDL}

To calculate fluctuations of the number of atoms in the
condensate, $n_{0}$, we need to find the expectation value of $n_{0}^{p}$ with
$p\leqslant2$, i.e. the first and second moments of the distribution. To find
these, we again use a generating function approach, introducing
\begin{equation}
\mathcal{G}_{p}(u)=%
{\displaystyle\sum\limits_{n=0}^{\infty}}
\left\langle n_{0}^{p}\right\rangle _{n}^{\text{can}}u^{n}\mathbb{Z}_{n}.
\end{equation}
Here, $\left\langle ...\right\rangle _{N}^{\text{can}}$ denotes the
expectation value in the canonical ensemble for $N$ particles. This evaluates
to $\mathcal{G}_{p}(u)=f_{p}(ub)\mathcal{G}_{0}(u)$ with
\begin{equation}
f_{p}(ub)=(1-ub)\sum_{m=0}^{\infty}m^{p}\left(  ub\right)  ^{m}.%
\end{equation}
For $p=1,2$ this simplifies to
\begin{align}
\mathcal{G}_{1}(u) &  =\frac{ub}{1-ub}\mathcal{G}_{0}(u),\\
\mathcal{G}_{2}(u) &  =\frac{ub(1+ub)}{(1-ub)^{2}}\mathcal{G}_{0}(u).
\end{align}
The expectation values are then extracted from the generating functions by
applying the same technique as outlined in \cite{BrosensEtAl1997a} for the partition
functions, namely through writing out%
\begin{equation}
\left\langle n_{0}^{p}\right\rangle _{N}^{\text{can}}=\frac{1}{\mathbb{Z}_{N}%
}\frac{1}{N!}\left.  \frac{d^{N}\mathcal{G}_{p}(u)}{du^{N}}\right\vert _{u=0}%
\end{equation}
by recursively applying the derivative with respect to $u$. This yields%
\begin{equation}
\left.  \frac{d^{N}\mathcal{G}_{p}(u)}{du^{N}}\right\vert _{u=0}=\sum_{\ell
=0}^{N}\frac{N!}{\ell!}\left.  \frac{d^{\ell}\left[  f_{p}(ub)\right]
}{du^{\ell}}\right\vert _{u=0}\mathbb{Z}_{N-\ell},%
\end{equation}
from which
\begin{equation}
\left\langle n_{0}^{p}\right\rangle _{N}^{\text{can}}=\sum_{\ell=0}^{N}%
\frac{1}{\ell!}\left.  \frac{d^{\ell}\left[  f_{p}(ub)\right]  }{du^{\ell}%
}\right\vert _{u=0}\frac{\mathbb{Z}_{N-\ell}}{\mathbb{Z}_{N}}.%
\end{equation}
In particular, for $p=1$ and $2$ we obtain%
\begin{equation}
\left\langle n_{0}\right\rangle _{N}^{\text{can}}=%
{\displaystyle\sum\limits_{\ell=1}^{N}}
\dfrac{b^{\ell}\mathbb{Z}_{N-\ell}(w)}{\mathbb{Z}_{N}(w)},%
\end{equation}
and%
\begin{equation}
\left\langle n_{0}^{2}\right\rangle _{N}^{\text{can}}=%
{\displaystyle\sum\limits_{\ell=0}^{N}}
\dfrac{(2\ell-1)_{+}b^{\ell}\mathbb{Z}_{N-\ell}}{\mathbb{Z}_{N}},%
\end{equation}
where $(a)_{+}=\max[a,0]$. From these moments we get the condensate
fluctuations through
\begin{equation}
\left(  \Delta n_{0}\right)  ^{\text{can}}=\sqrt{\left\langle n_{0}%
^{2}\right\rangle _{N}^{{\text{can}}}-\left(  \left\langle n_{0}\right\rangle
_{N}^{{\text{can}}}\right)  ^{2}}.%
\end{equation}

\end{appendix}

%

\end{document}